\documentclass[aps,prd,twocolumn,superscriptaddress,showpacs,nofootinbib]{revtex4}

\usepackage{color}
\usepackage{graphicx}
\usepackage{amsmath,amssymb,latexsym}
\usepackage{bm}
\usepackage{epsfig}
\newcommand{\postscript}[2]{\setlength{\epsfxsize}{#2\hsize}
   \centerline{\epsfbox{#1}}}

\begin{document}

\title{Present and Future Gamma-Ray Probes of the Cygnus OB2
  Environment}

\author{Luis A.~Anchordoqui}
\affiliation{Institut de Ci\`encies de l'Espai (IEEC-CSIC),
Campus UAB, Torre C5, 2a planta, 08193 Barcelona, Spain}
\affiliation{Department of Physics, University of Wisconsin-Milwaukee,
P.O. Box 413, Milwaukee, WI 53201, USA
}

\author{Haim Goldberg}
\affiliation{Department of Physics, Northeastern University, Boston,
  MA 02115, USA
}

\author{Russell D.~Moore}
\affiliation{Department of Physics, University of Wisconsin-Milwaukee,
P.O. Box 413, Milwaukee, WI 53201, USA
}

\author{Sergio Palomares-Ruiz}
\affiliation{Centro de F\'{\i}sica Te\'orica de Part\'{\i}culas,
Instituto Superior T\'ecnico, 1049-001 Lisboa, Portugal}

\author{Diego F.~Torres}
\affiliation{Institut de Ci\`encies de l'Espai (IEEC-CSIC), Campus
  UAB, Torre C5, 2a planta, 08193 Barcelona, Spain}
\affiliation{Instituci\'o Catalana de Recerca i Estudis Avan\c{c}ats
  (ICREA), Spain}

\author{Thomas J.~Weiler}
\affiliation{Department of Physics and Astronomy, Vanderbilt
  University, Nashville, TN 37235, USA
}

\date{\today}

\begin{abstract}
\noindent
The MAGIC Collaboration has provided new observational data pertaining
to the TeV~J2032+4130 gamma-ray source (within the Cygnus~OB2 region),
for energies $E_\gamma >400$~GeV. It is then appropriate to update the
impact of these data on gamma-ray production mechanisms in stellar
associations. We consider two mechanisms of gamma-ray emission, pion
production and decay (PION) and photo-excitation of high-energy nuclei
followed by prompt photo-emission from the daughter nuclei
($A^\star$).  We find that while the data can be accommodated with
either scenario, the $A^\star$ features a spectral bump, corresponding
to the threshold for exciting the Giant Dipole Resonance, which can
serve to discriminate between them. We comment on neutrino emission
and detection from the region if the PION and/or $A^\star$ processes
are operative. We also touch on the implications for this analysis of
future Fermi and \v Cerenkov Telescope Array data.
\end{abstract}

\pacs{07.85.-m, 98.70.Sa
\hfill CFTP/09-27}
\maketitle

Two well-known mechanisms for generating TeV $\gamma$-rays in
astrophysical sources are the purely electromagnetic (EM) one
proceeding via synchrotron emission and inverse Compton scattering,
and the hadronic (PION) one in which $\gamma$-rays originate from
$\pi^0$ production and decay~\cite{S71,Aharonian:2004yt}. Recently, we
highlighted a third dynamic which leads to TeV $\gamma$-rays:
photo-excitation of high-energy nuclei, followed by prompt
photo-emission from the excited daughter
nuclei~\cite{Anchordoqui:2006pd,Moskalenko}. In this chain reaction,
the nuclei act in analogy to Einstein's relativistic moving mirror to
``double-boost'' eV~starlight to TeV energies for a Lorentz boost
factor $\agt 10^6$. The important role played by the Giant Dipole
Resonance (GDR) in the photo-disintegration effectively suppresses the
contribution to the $\gamma$-ray spectrum below a
TeV~\cite{Ioka:2009dh}. This process (which we have denoted by
$A^\star$) has been proposed~\cite{Anchordoqui:2006pe} as a candidate
explanation of the unidentified HEGRA source
TeV~J2032+4130~\cite{Aharonian:2002ij} at the edge of the Cygnus~OB2
(Cyg~OB2) association (see Ref.~\cite{W2} for the discussion regarding
the HESS source HESS~J1023-575 at the edge of
Westerlund~2~\cite{HESSW2}). This stellar association has been known
to harbor a large population of massive and early type
stars~\cite{K00} that can provide the required UV target density.

Recently, the MAGIC Collaboration has reported new TeV $\gamma$-ray
data from this region~\cite{Albert:2008yk}. Thus, it is of interest
to expand previous analyses to include this new data. In this paper,
we present a unified analysis of the $A^\star$ and PION mechanisms for
gamma-ray production in Cyg~OB2.  This combined analysis indicates the
relative importance of these two mechanisms. Extrapolation to lower
energies then allows us to make predictions within reach of the Fermi
mission~\cite{Atwood:2009ez}.

TeV gamma-ray data from this region has also been reported by the
Milagro Collaboration~\cite{MILAGRO}. However, inclusion of these data
in our study will require further analysis to distinguish
contributions from the HEGRA source and diffuse interactions expected
from the larger region of observation. Thus we postpone the
consideration of the Milagro data.

The critical parameters for the PION and $A^\star$ mechanisms are the
ambient hydrogen gas density and the UV photon background,
respectively. These parameters at present are subject to considerable
uncertainty. It is a further goal of this work to see whether the
present and evolving gamma-ray data can meaningfully constrain these
parameters.

The stellar distribution of Cyg~OB2 reveals a rather regular and
almost circular density profile with the center located at $(\alpha,
\delta) = (20^h 33^m 10^s, +41^o 12')$ and with a pronounced maximum
slightly offset at $(\alpha, \delta) = (20^h 33^m 10^s, +41^o
15.7')$~\cite{K00}. Star counts show that 50\% of the members are
located within a radius of $21'$, and 90\% within a radius of $45'$
around the center. By integrating the radial density profile, after
subtraction of the field star density, the total number of OB stars
is found to be $2600 \pm 400$, with a O-type star population of $120
\pm 20$. This suggests that the total mass of the association is about
$10^{4} M_\odot$. Distance determinations set the proximity of the
Cyg~OB2 to $d \sim 1.7~{\rm kpc}$~\cite{TTC91}. At such distance, the
inner $21'$, with half the total number of objects, results in a
physical radius of $R_{\rm in} \sim 10~{\rm pc}$, with $R_{\rm out}
\sim $~30~pc being the radius of the association. Projected onto the
sky at the distance of Cyg OB2, the HEGRA/MAGIC signal from
TeV~J2032+4130~\cite{Aharonian:2002ij,Albert:2008yk} was observed only
in a 3~pc radius cell at the edge of the association.  With the same
angular radius, there are a total of $\sim 14$ cells in the core of
the association.  The flux in each of these cells is bounded $\sim1\%$
of Crab, about 3 times less than that of the TeV~J2032+4130 cell. On
the other hand, the age of the stellar association is supposed to be
2-4~Myr~\cite{age}.

As mentioned above, the prediction of the PION $\gamma$-ray yield is
subject to uncertainty in the ambient gas density (as well as the
ambient cosmic ray flux). We will find that for gas densities larger
than $0.1~{\rm cm}^{-3}$ in the vicinity of the source, the PION
mechanism will dominate, with the $A^\star$ mechanism assuming
dominance for smaller densities. A considerably higher density,
$n_{\rm H} = 30~{\rm cm^{-3}}$, has been estimated from observations
of the CO $J=1 \to 0$ rotational transition~\cite{Butt:2003xc}. If
taken as representing an average over the core of the association,
this value implies a hydrogen gas mass of 3000~$M_\odot,$ which is
about 30\% of the mass found in stars. Arguments have been
given~\cite{Torres:2003ur} that this estimated
density~\cite{Butt:2003xc} should be interpreted as an upper
bound. Among concerns are the CO-H$_2$ conversion
factor~\cite{Yao:2003rm}, and the size of the region used for
averaging.  More recently, a thorough analysis of the region, with
higher angular resolution observations of $^{12}$CO and $^{13}$CO has
been presented~\cite{Butt:2008}. Of interest in these results is a
significant $^{13}$CO cavity right at the TeV source position.  This
is reminiscent of a formerly found IR void. Only one  $^{13}$CO clump
is seen at a position consistent with the projected position of HEGRA
source. No claim is made as to any physical connection between the
clump and the cavity. Although the clump is massive (claimed mass at
337~M$_\odot$), its size is significantly smaller than the size of the
extended TeV source. If gamma-rays were to originate from interactions
with just this clump, there is no reason for the source to appear
extended in instruments such as MAGIC~\cite{Cortina:2005pt} or
VERITAS~\cite{Holder:2006gi}.

We adopt as our fiducial density the low estimate $n_{\rm H} \sim
0.1~{\rm cm}^{-3}$, which agrees with the analysis
in Ref.~\cite{Torres:2003ur}.  This choice allows us to illustrate the
crossover point of PION dominance versus $A^\star$ dominance. It is
conservative and wise at this point to depend on future experiments
(discussed below) to resolve the issue of PION versus $A^\star$
dominance.

First we discuss the PION mechanism for TeV~gamma-ray production. The
emissivity (number/volume/time/energy) of neutral pions resulting from
an isotropic distribution of highly relativistic nuclei having a
power-law energy spectrum $dn_A (E_N)/dE_N = N_A (E_N/E_0)^{-\alpha}$,
colliding with ambient hydrogen, is given by~\cite{S71}
\begin{eqnarray}
\label{Qpi}
Q_{\pi^0}^{Ap} (E_{\pi^0}) & = &  c\,n_{\rm H} \, \int_{E_N^{\rm th}
  (E_{\pi^0})}^{E_N^{\rm max}} \frac{dn_A}{dE_N} (E_N) \nonumber \\
 & \times &
  \frac{d\sigma_A}{dE_{\pi^0}} (E_{\pi^0},E_N) \, dE_N
\end{eqnarray}
where $N_A$ is the normalization constant with units 1/volume/energy,
$E_0$ is set to 1~TeV, $E_N^{\rm th} (E_{\pi^0})$ is the minimum
energy per nucleon required to produced a pion with energy
$E_{\pi^0}$, and $d\sigma_A(E_{\pi^0},E_N)/dE_{\pi^0}$ is the
differential cross section for the production of a pion with energy
$E_{\pi^0}$ in the lab frame due a nucleus $A$ of energy per nucleon
$E_N=E_A/A$ colliding with a hydrogen atom at rest. The differential
cross-section can be parametrized by
\begin{equation}
\frac{d\sigma_A}{dE_{\pi^0}} (E_{\pi^0},E_N) \simeq
\frac{\sigma_0^A}{E_{\pi^0}} \, x F_{\pi^0}(x, E_N) \,,
\label{parametrization}
\end{equation}
where $x \equiv E_{\pi^0}/E_N$, $\sigma_0^A = A^{3/4} \, \sigma_0$
provides a scaling of the cross-section with the atomic
number~\cite{LST63}, $\sigma_0 = (34.3+1.88 L + 0.25 L^2)$~mb, and
$F_{\pi^0}(x, E_N)\equiv dN_{\pi^0}/dx$ is a fragmentation
function. We take
\begin{eqnarray}
F_{\pi^0} (x, E_N) & = & 4 \beta B_\pi x^{\beta -1} \,
  \left(\frac{1-x^{\beta}}{1+ r \, x^{\beta} \,
    (1-x^{\beta})}\right)^4 \nonumber \\
 & \times &
  \left(\frac{1}{1-x^{\beta}} + \frac{r \, (1-2 x^{\beta})}{1 + r \,
  x^{\beta} \, (1 - x^{\beta})}\right) \,,
\label{fpi}
\end{eqnarray}
with $B_\pi = a + 0.25$, $\beta = 0.98/\sqrt{a},$ $r= 2.6/\sqrt{a},$
$a = 3.67 + 0.83L + 0.075 L^2,$ and $L = \ln (E_N/{\rm
  TeV})$~\cite{Kelner:2006tc}. Because isotropy is implied in
(\ref{Qpi}), it is straightforward to obtain the $\gamma$-ray
emissivity~\cite{Anchordoqui:2006pe}; it is
\begin{eqnarray}
Q_{\gamma}^{Ap} (E_\gamma) & = & 2 \, \int_{E_{\pi^0}^{\rm min}
  (E_\gamma)}^{E_{\pi^0}^{\rm max} (E_{N}^{\rm max})} \, \frac{Q_{\pi^0}^{Ap}
  (E_{\pi^0})}{\left(E_{\pi^0}^2 - m_{\pi}^2 \right)^{1/2}} \,
  dE_{\pi^0} \,
\label{iiioi}
\end{eqnarray}
where $E_{\pi^0}^{\rm min} (E_\gamma) = E_\gamma +
m_{\pi}^2/(4E_\gamma)$.

Before proceeding to the $A^\star$ mechanism, we pause to compare
Eq.~(\ref{fpi}), which is a functional fit to the outcome of numerical
simulations obtained with the SIBYLL event
generator~\cite{Fletcher:1994bd}, to data collected at Tevatron by the
CDF detector~\cite{Abe:1989td}.  The results of simulations leading to
Eq.~(\ref{fpi}) have been reported at nucleon energies of 0.1~TeV and
1000~TeV.  The latter corresponds to a center-of-mass energy of
$\sqrt{s} \simeq 1410~{\rm GeV}$. The CDF group at Tevatron has
measured the charged pion spectrum for pseudorapidity $|\eta| < 3.5,$
at cm energies of 630~GeV and 1800~GeV. They have provided a fit over
the energy range of interest, quadratic in $\ln [s/{\rm GeV}^2]$, with
$\chi^2 = 0.72$ for three degrees of freedom
\begin{eqnarray}
\frac{dN_{\rm ch}}{d \eta} & = & (0.023 \pm 0.008) \ln^2 s - (0.25 \pm
0.19)
\ln s \nonumber \\
& + &  (2.5 \pm 1.0) \, ,
\end{eqnarray}
valid for $\eta =0$. Taking into account that $F_{\pi^+} \simeq
F_{\pi^-} \simeq F_{\pi^0}$, that the spectral dependence on $\eta$ is
mild, and that
\begin{equation}
\frac{dN_{\pi^0}}{d \eta} \simeq x \frac{dN_{\pi^0}}{dx} = x F_{\pi^0} \,,
\label{multiplicity}
\end{equation}
we find that the results in Eq.~(\ref{fpi}) agree remarkably, within one
standard deviation, to the CDF fit to their Tevatron data.

Now we discuss the $A^\star$ mechanism of TeV~gamma-ray production.
The photo-excitation (or photo-disintegration) rate for a highly
relativistic nucleus with energy $E=\gamma A m_N$ (where $\gamma$ is
the Lorentz factor) propagating through an isotropic photon background
with energy $\epsilon$ and number-density spectrum $n(\epsilon)$
is~\cite{S69}
\begin{equation}
R_A  =
\frac{1}{2} \, \int_0^{\infty} \frac{n(\epsilon)}{\gamma^2 \epsilon^2}
\, d\epsilon \, \int_0^{2\gamma \epsilon} \epsilon' \,
\sigma_A(\epsilon') \, d\epsilon' \, ,
\label{rate}
\end{equation}
where $\sigma_A (\epsilon')$ is the cross section for
photo-disintegration of a nucleus of mass $A$ by a photon of energy
$\epsilon'$ in the rest frame of the nucleus.

We assume that $n(\epsilon)$ results from thermal emission of  the
stars in the whole Cyg~OB2 association, out to $R_{\rm out}\sim$~30~pc.
We model the association with half of the stars uniformly distributed
in the inner region, $R_{\rm in} \sim$~10~pc, and the other half
uniformly distributed in the outer shell, i.e.\ the density of stars
in the inner region is $(R_{\rm out}/R_{\rm in})^3 - 1 \sim 26$ times
that in the outer shell. To reproduce the size and position of the
source of the HEGRA signal, the photo-disintegration must occur in a
region of radius $r \sim 3$~pc at the edge of the inner part of the
association, $R\le R_{\rm in}$.  The average photon density in this
region reflects both the temperatures $T_{\rm O}$ and $T_{\rm B}$ of
the O and B stars, respectively, and dilution resulting from inverse
square law considerations. The resulting photon density is
\begin{equation}
n^{\star}(\epsilon) = \frac{47}{4} \ \left[\frac{n_{\rm O}(\epsilon)
    \,\, N_{\rm O} \,R_{\rm O}^2 + n_{\rm B}(\epsilon) \,\, N_{\rm B}
    \,\,R_{\rm B}^2}{R_{\rm out}^2} \right] \,\,,
\end{equation}
where $N_{\rm O\,(B)}$ is the number of O~(B) stars , $R_{\rm O(B)}$
is the O~(B) star average radius, and
\begin{equation}
n_{\rm O(B)} (\epsilon) = (\epsilon/\pi)^2\
\left[e^{\epsilon/T_{\rm O(B)}}-1 \right]^{-1} \,,
\label{nBE}
\end{equation}
is the Bose-Einstein distribution of photons emitted from a star at
temperature $T_{\rm  O(B)}$. The factor $47/4$ is a consequence of
averaging the inverse square distance within this distribution for the
density and the region where the reaction takes
place~\cite{Anchordoqui:2006pe}.  It is clear, however, that within
the 3~pc HEGRA hot spot the concentration of stars would be above
average, and thus hereafter we take as a fiducial value for $n^{\rm
  HEGRA}(\epsilon) = 1.7 n^\star (\epsilon)$. The 1.7 factor
encapsulates an uncertainty of $\sim 1$~to~2.5~\cite{nhegra}. The
resulting photo-disintegration rate $R_A$ for the value of this density
will be denoted by $R_A^{\rm HEGRA}$.

\begin{figure}
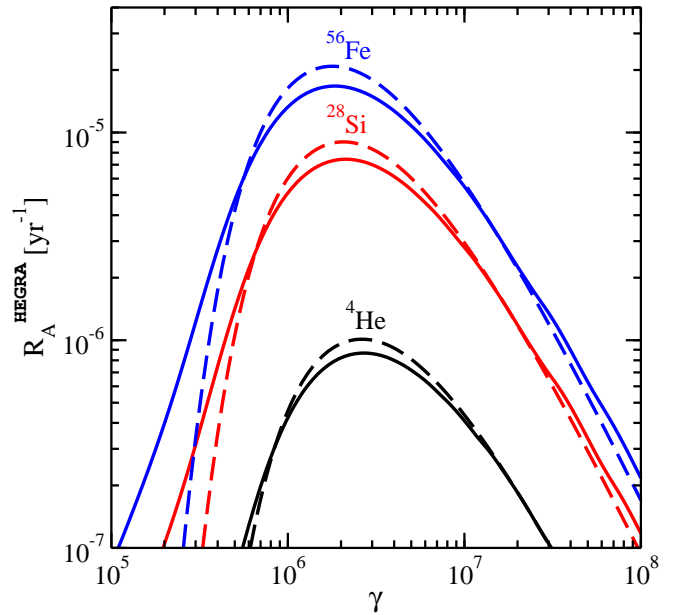

\begin{center}
\postscript{RstarsBW17.eps}{1}
\end{center}
\caption{Photo-disintegration rates of $^{56}$Fe, $^{28}$Si, and
  $^{4}$He for the HEGRA hot spot. We have approximated the cross
  section of the GDR by a dipole (solid-lines) and by a single pole of
  the NWA (dash-lines).}
\label{Rstar}
\end{figure}

In Fig.~\ref{Rstar} we show the dependence on the Lorentz factor of
$R_{A}^{\rm HEGRA}$, for the stellar ambiance described above. For the
O stars we have taken $N_{\rm O} = 130$, a surface temperature $T_O$ =
40000~K, and radius $R_O = 19~R_\odot$; for the cooler B stars we
assign $T_B$ = 18000~K, $N_{\rm B} = 2470$, and radius $R_B =
8~R_\odot$.  The numbers $N_{\rm O}$ and $N_{\rm B}$ are consistent
with the Cyg~OB2 data discussed in the introduction. The cross section
has been calculated in both the Narrow-Width Approximation (NWA) and
the more accurate dipole approximation, for the nuclear parameters
given in Ref.~\cite{Anchordoqui:2006pe}. In the calculation that
follows we adopt the more accurate dipole form for the cross section.

The low-energy cutoff on $R_A^{\rm HEGRA}$ is evident in
Fig.~\ref{Rstar}. This cutoff will be mirrored in the resulting
gamma-ray distribution. Notice that the NWA, which we do not use
below, overestimates the severity of the low-energy cutoff.

The energy behavior for photons in the $0.5-10$~TeV region of the
HEGRA and MAGIC data is a complex convolution of the energy
distributions of the various nuclei participating in the
photo-disintegration, with the rate factors appropriate to the eV
photon density for the various stellar populations. Approximating the
$\gamma$-ray spectrum as being monochromatic with energy
$\overline{E'_{\gamma A}}$ (in the nucleus rest frame), the emissivity
becomes~\cite{Anchordoqui:2006pe}
\begin{eqnarray}
\label{qgdis}
Q_\gamma^{A^\star}(E_\gamma) & = & \sum_A \frac{\overline{N_A} m_N}{2
  \overline{E'_{\gamma A}}} \int_{\frac{m_N E_{\gamma
  }}{2\overline{E'_{\gamma A}}}}   \frac{dE_N}{E_N}
 \nonumber \\
 & \times & R_A^{\rm HEGRA} (E_N) \, \frac{dn_A}{dE_N}(E_N)\, ,
\end{eqnarray}
where  $E_\gamma$ is the energy of the emitted $\gamma$-ray in the
lab, and $\overline{N_A}$, which we take to be 2~\cite{Fe}, is the
mean $\gamma$-ray multiplicity for a nucleus with atomic number $A$.

It is important to note that the same nucleus source density $dn_A
/dE_N$ is present in the $A^\star$ emissivity  (\ref{qgdis}) and in
the PION emissivity (\ref{iiioi}) (via (\ref{Qpi})). Thus, a
comparison of the two mechanisms will depend only weakly on the exact
features of $dn_A /dE_N$.

\begin{figure}
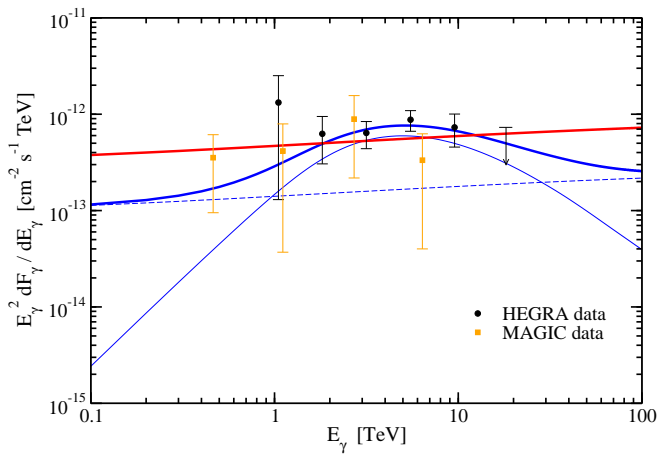

\begin{center}
\postscript{HME2_APION_2_nocutoff.eps}{1}
\end{center}
\caption{Eyeball fits to HEGRA and MAGIC $\gamma$-ray spectrum. We
  have assumed an iron nuclei population with $\alpha = 2$ and
  $\overline{E'_{\gamma \rm Fe}}= 2~{\rm MeV}$. The thick straight red
  line is a representative fit to the combined spectra assuming the PION
  process only with $n_{\rm H} = 2~{\rm cm}^{-3}$ and $N_{\rm Fe} = 5
  \times 10^{-12}~{\rm cm}^{-3}\, {\rm TeV}^{-1}$. The thick solid blue
  curve is a similar fit combining both the $A^*$ mechanism (thin
  solid blue line) and PION mechanism (dashed blue line) for $n_{\rm
    H} = 0.1~{\rm cm}^{-3}$ and $N_{\rm Fe} = 3 \times 10^{-11}~{\rm
    cm}^{-3}\, {\rm TeV}^{-1}$.}
\label{specs}
\end{figure}

The differential photon flux at the observer's site (assuming there is
no absorption) receives contributions from both mechanisms, PION and
$A^\star$. The result is related to to the $\gamma$-ray emissivity as
\begin{equation}
\label{fluxg}
\frac{dF_\gamma}{dE_\gamma} (E_\gamma) =
\frac{V_{\rm dis}}{4 \pi d^2} \, [Q_{\gamma}^{Ap} (E_\gamma)  +
Q_\gamma^{A^\star} (E_\gamma)]  \,,
\end{equation}
where $V_{\rm dis}$ is the volume of the source region and $d$ is the
distance to the observer. In Fig.~\ref{specs} we provide a some
eyeball fits (thick solid lines) to the combined HEGRA/MAGIC
$\gamma$-ray spectrum, obtained from integrations implicit in the two
emissivities in Eq.~(\ref{fluxg}).

The fits are for an iron nuclei population with spectral index $\alpha
= 2$ and for an average energy of the photon (in the nuclear rest frame)
emitted during photo-emission $\overline{E'_{\gamma \rm Fe}}= 2~{\rm
  MeV}$~\cite{Fe}. The solid thick blue curve is a fit using both the
$A^*$ mechanism (solid blue thin line) and PION mechanism (dash blue
thin line), with  $n_{\rm H}$ equal to our fiducial value, $0.1~{\rm
  cm}^{-3}$, and $N_{\rm Fe} = 3 \times 10^{-11}~{\rm cm}^{-3}\, {\rm
  TeV}^{-1}$. The red thick straight line is a representative fit to
the combined spectral data assuming the PION process only with $n_{\rm
  H} = 2~{\rm cm}^{-3}$ and $N_{\rm Fe} = 5 \times 10^{-12}~{\rm
  cm}^{-3}\, {\rm TeV}^{-1}$. For the iron nuclei population assumed
in the fits, the target gas density for PION dominance at all energies
is $n_{\rm H} \gtrsim 0.5~{\rm cm}^{-3}$.

Additional data is becoming available from observations of the Fermi
satellite. A preliminary measurement in the Cygnus region yields an
integrated gamma-ray flux~\cite{Abdo:2009mg}
\begin{equation}
  F_\gamma (1-100~{\rm GeV}) \simeq 3.07 \times 10^{-8}~{\rm cm}^{-2}
  \, {\rm s}^{-1} \, .
\end{equation}
If one na\"ively assumes a spectrum $\propto E_\gamma^{-2}$ for their
observations, a squared-energy weighted differential flux of
\begin{equation}
E_\gamma^2 \frac{dF_\gamma}{dE_\gamma} = 3.07 \times 10^{-11} {\rm
  cm}^{-2} \, {\rm s}^{-1} \, {\rm TeV}
\end{equation}
is obtained. This is nearly two orders of magnitude above
extrapolation of the HEGRA/MAGIC measurement at 100~GeV. The presence
of low energy powerful sources can clearly dominate the flux at lower
energies~\cite{Hartman:1999fc}.  Thus, the normalization inferred from
this low-energy data can grossly over-estimate the predicted flux at
100~GeV (see Ref.~\cite{Funk:2007zzd} for a thorough discussion of
these issues).  A prime candidate source for the low energy radiation
is a pulsar (with spin-down power $2.6 \times 10^{35} {\rm erg} {\rm
  s}^{-1}$), which coincides (within errors) with the position of
TeV~J2032+4130 ($4'$ displacement)~\cite{fermilat}.  The observed GeV
emission can plausibly be ascribed to electron acceleration in the
magnetosphere. This radiation is exponentially cut off in the TeV
region. However, such very high energy radiation, which is the focus
of the present paper, could possibly be associated with inverse
Compton scattering of the electrons which power the pulsar wind
nebula, if such exists (see as an example the case of HESS
J1825-137~\cite{HESSJ1825-137}, and many others in the recent
literature). Alternatively, the TeV radiation can originate in the OB
association via the $A^\star$ and PION mechanisms.  \bigskip

We now conclude with a discussion of our results:

\begin{itemize}

\item From Fig.~\ref{specs} it is apparent that the combined
  HEGRA/MAGIC data can be fit with only the PION mechanism in
  operation. Such a fit applies if the gas density $n_H$ is larger
  than $2~{\rm cm}^{-3}$.  For $0.05~{\rm cm}^{-3} \lesssim n_H
  \lesssim 2~{\rm cm}^{-3}$, a combination of PION and $A^\star$ can
  provide a satisfactory fit to the data, whereas for $n_{\rm H} <
  0.05~{\rm cm}^{-3}$ a good fit to all the data can be obtained using
  only the $A^\star$ mechanism.

\item The average energy of the photon (in the nucleus rest frame)
  emitted during photo-emission has been taken as 2~MeV. This is
  appropriate for iron nuclei. If $n_{\rm H} < 0.05~{\rm cm}^{-3}$, a
  better fit to all the data can be obtained using only the $A^\star$
  mechanism with a lower average energy of 1.5~MeV.

\item For low gas densities, the spectral features characteristic of
  the $A^\star$ mechanism become visible. These are best described as
  a broad bump in the spectrum in the region $1-10$~TeV.

\item In completing the explanation of the HEGRA and MAGIC signal,
  there remains one issue to address -- the signal was observed only
  in a 3~pc radius cell at the edge of the inner association. At this
  point in our understanding we can provide only qualitative remarks.
  One possibility is an increased density of very hot OB stars in the
  TeV~J2032+4130 cell, which provide efficient trapping and
  accelerating conditions  for the nuclei, as well as a hot photon
  background. Indeed, a recent estimate~\cite{Butt:2005dx} indicates
  around 10~O stars in the region of the source, a number which is a
  factor of 3 larger than that expected on the basis of a uniform
  population.

\item If the energy spectrum of cosmic electrons $\propto E_e^{-2}$
  (with an exponential cutoff at 40~TeV), the data can also be
  explained by inverse Compton scattering of these electrons on the
  cosmic microwave background photons~\cite{Albert:2008yk}. The EM
  explanation can only accommodate the data if the Compton peak is
  matched to the energy range of HEGRA/MAGIC detection, a possibility
  allowed within errors.

\item We expect a flux of TeV $\nu_\mu, \bar\nu_\mu$ from both the
  $A^\star$ (via neutron decay followed by
  oscillations~\cite{Anchordoqui:2003vc}) and PION (via $\pi^\pm$
  decay)~\cite{Torres:2003ur} mechanisms.  Allowing about one muon
  neutrino per photon after oscillation, we expect about 1.2~events/yr
  at IceCube with a background from atmospheric neutrinos of about 1
  event/yr~\cite{Anchordoqui:2005gj}. However, it is possible that
  this event rate can be considerably enhanced by emission from the
  additional 3-pc cells in the association (which will not be resolved
  by future neutrino detectors). The signal enhancement can amount to
  as much as a factor of about 5 due to the emission at the upper
  limit value set by gamma-ray observation from each cell in the rest
  of region (e.g., MILAGRO measurement in a region centered in the
  HEGRA region but ten times larger~\cite{MILAGRO} and MAGIC upper
  limit in the direction of Cyg~X3~\cite{Albert:2008yk}, which
  approximately coincides with that of Cyg~OB2). Such accumulation
  could make the source visible in neutrinos at IceCube. We also note
  that absorption of gamma-rays at the center of the association (see
  e.g., Ref.~\cite{DomingoSantamaria:2005yh}) could be relevant,
  implying an even higher neutrino flux from some cells. Observation
  of a neutrino flux from the HEGRA/MAGIC source could disqualify an
  EM explanation of the origin of the gamma-rays, at least for this
  source.

\item The future \v{C}erenkov Telescope Array~\cite{Hermann:2007ve}
  will provide stronger spectral discrimination between the PION and
  PION+$A^\star$ mechanisms. This telescope is projected to have a
  factor $>10$ larger sensitivity than MAGIC/VERITAS at TeV
  energies. It will also cover the lower GeV energy region (down to
  tenths of a GEV) where the $A^\star$ mechanism is suppressed, thus
  allowing the possibility of comparing the two mechanisms with a
  single data set covering the entire energy region of interest. Due
  to its superb angular resolution (expected perhaps at a factor of 2
  or 3 better than that of MAGIC) and field of view (several degrees),
  it will become the ideal instrument to distinguish emission
  components in this energy region, and to study morphology of the
  radiation from TeV~J2032+4130.

\end{itemize}

\acknowledgments{
We thank Jordi Isern for a valuable communication. LAA is
supported by the US National Science Foundation (NSF) Grant No
PHY-0757598, the UWM Research Growth Initiative, and Consejo Superior
de Investigaciones Cient\'{\i}ficas (CSIC). HG is supported by the US
NSF Grant No PHY-0757959. SPR is partially supported by the Portuguese
FCT through CERN/FP/83503/2008 and CFTP-FCT UNIT 777, which are
partially funded through POCTI (FEDER), and by the Spanish Grant
FPA2008-02878 of the MCT. DFT is supported by Spanish Grants
AYA2006-00530 and AYA2008-01181-E/ESP. TJW was supported by the US
Department of Energy (DoE) Grant DE-FG05-85ER40226, an Alexander von
Humboldt Foundation Senior Research Award, the faculty leave program
of Vanderbilt University, and the hospitality of the Technische
Universit\"at Dortmund, and the Max-Planck-Instituts f\"ur Physik
(Heisenberg-Institut), M\"unchen, and f\"ur Kernphysik,
Heidelberg. HG, SPR, and TJW thank the Aspen Center for Physics where
this paper was finished.
}


\begin{thebibliography}{99}
\bibitem{age}
  The superposition of the isochrones (calculated using the theoretical
  evolutionary tracks~\cite{isochrone}) on the Hertzsprung-Russell
  diagrams, suggests the age of the association is 1--4 Myr~\cite{K02}.
  This range reflects the dispersion of the upper main sequence and
  agrees with the fact that the observed large number of O-type stars
  implies that the association should be younger than $\sim$5~Myr,
  because in the case of coeval star formation, the number of this type
  of stars decreases rapidly. In addition, the presence of some
  Wolf-Rayet stars within Cyg OB2 implies an age larger than $\sim
  2~{\rm Myr}$, while the non-detection of any supernova
  remnant~\cite{WHL91} points to an association younger than $\sim
  4~{\rm Myr}$.
\bibitem{nhegra}
  This choice of $n^{\rm HEGRA} (\epsilon)$ agrees with an estimate in
  which the densities at the sources are diminished by a common factor
  $\langle {1/r_{ij}}^2 \rangle$, where the latter is simply the
  average value of the inverse square of the star spacings, taken over
  the core of the association~\cite{Anchordoqui:2006pe}. This is an
  overly conservative estimate of the fluctuation: if about 10~\% of
  the O and B stars are within the 3~pc radius of the HEGRA source,
  the UV photon population in this region due to emission from {\em
  these stars alone} can be calculated to be 2.5 $n^\star(\epsilon)$.
\end{thebibliography}
\end{document}